# Levy's distributions of the currents intensities in metal-dielectric systems

M. Mokhtari*, L. Zekri**

USTO, Département de Physique, LEPM, BP 1505 El M'Naouar, Oran, Algérie

E-mail :   *mohamed.mokhtari@yahoo.fr
\*\* lzekri@yahoo.com

**Abstract**: Levy distributions are involved in many physical phenomenons. The purpose of our study is to understand metal-insulator transitions in composite systems using the above distributions. The exact method (EM) based on the calculation of Kirchhoff's equations solutions is used to compute currents intensities in each network division. We study currents intensities distributions for different widths of the system. We interpret these distributions in terms of Levy's law.

**Introduction**

Percolation theory is an old model for disordered systems and its phase's transitions [1]. Different study methods and technic's are used, analytically (effective medium theory, field theory…etc) and numerically (random resistors network, Monte Carlo simulation…) [2].

The construction of continuous macroscopic objects (aggregates, piles) using the random spreading of particles or links had been the subject of many studies during last decade, and been formalized by divers percolation models [1, 2, 3, 4]. The main amounts are expressed with power's degrees laws close to critical points, and the main results are resumed by a list of critical exponents, which depend only on dimension and not on models details [1]. The increment between 2D and 3D have been subject of many studies, for instance, in magnetic phenomena, it's used to describe phase transition behavior of thin magnetic films in the case of magnetic phenomena [5]. For studying conductivity of thin composite systems, clerc and al considered the increment between 2D and 3D [6]. More recently, Zekri et al [15] investigated the effect of thickness on the percolation threshold and conductivity exponent of metal-insulator composites.[15]

In the present paper, we focus on the thickness dependence of the percolation threshold as well as the fluctuation width for thin and thick films corresponding roughly to 2D and 3D systems using the Levy distribution.

**Method description:**

We consider a random network of resistors, of *NxNxh* cubic elementary cells, *h* is the system's thickness and *N* is the size of the perpendicular plane (Fig. 1). The network is uniformly and randomly filled with metallic and insulating components, having conductivities of    = 1   and    = 10  with filling densities of    and 1 −   . One   volt dc is applied in the two bounds of the network, as specified in fig1.

A various calculation methods are used to numerically define the networks' conductivity. Amongst them we mention approximation methods, like the renormalization method (RSRG) [14], exact transformations methods, like Lobb and Frank method [13] and transfer matrix method [12]. These methods are limited to 2D or shows numerical instability for large systems. We use the exact method in our work [9].  which is based on the numerical resolution of Kirchhoff's equations in network, leading to effective conductance determination (admittance or impedance ) and current or local field distribution in each component (or network link) as well.

In order to further understand the behavior of the increment between 2D and 3D, we investigate the current intensities distribution        for different system's thicknesses. Therefore, we use the Levy distributions

We consider a finite system characterized by    . For instance,    represent currents intensities in meshes. In one hand, the distribution (or the probability)        can leads to a low

-

where all events are distributed around a mean value, In this case, the usual central theorem is applied. The distribution form converges to a Gauss law. In the other hand,       can lead to a large law. In this case       converge to a Levy distribution, which is different than Gauss law, and was described by Levy and Gnedenko's generalized central limit theorem [7,11]. The infrequent events of "large" distribution (which forms a long "tail") break down smoothly like a power law shows:

$$( )_{\to \infty}$$

The length of the tail increases while µ decreases. We can consider some cases:

For    > 2 the square average of is finite and the central limit theorem is applied for large N the mean value of    converge. Nevertheless, this convergence can be very slow if    is close to 2.

For    < 1, note that <    > is infinite in this case. An intermediate status exists 1 <    < 2, then the term   , with    → ∞, represent an important part of the sum, and <    > is finite. Tail exponent can easily calculated owing to a histogram representation of        , the decay of which we fit by a power law

**Results and discussions:**

The computation of Currents intensities was performed by exact method (EM) based on the resolution of Kirchhoff's equations [8,9].. The Calculations are performed for various height and many layers. In fig.2 intensity distribution is shown around percolation threshold for thickness *h=15* . At the percolation threshold, two branches appear, one seems to be log normal for the great currents intensities, and the other for the low intensities. Below the threshold, the log-normal branch tends to disappear because fewer metal bonds build the backbone, above this threshold the second branch tends to disappear when all the bonds take part in the infinite cluster. The behavior shown in figure 2 is similar to that observed in red bonds  distribution show by zekri and al [10] (links crossed by the maximum current) Thus near the percolation threshold the largest cluster is composed mainly of critical links, their number is proportional to the conductivity. These links are responsible on the percolation-non-percolation transition.

Fig.3 shows the evolution of the exponent µ of tail when p vary. We can divide this evolution in three different parts:

- For µ>2  the two first moments <    > and <    $^2$ >   are finite, the distribution of       tends to a Gauss law when       → ∞ . Currents intensities are very strong and metallic links are high enough to rule out the total summation.

- For 1 <    < 2 , <    > is finite and <    $^2$ > is infinite, then we can't find percolation threshold in this domain.

-

- For  < 1 the two first moments <  > and <  ² > are infinite. This gives evidence that all moments diverge. In this zone the minimum of the exponent µ is who corresponds to the maximum of fluctuations of current and the threshold of percolation pc corresponds to the maximum of fluctuations. Which leads us to say that the threshold of percolation corresponds at least of parameter µ, one can get percolation threshold in this domain. For instance, for a size 100 and  thickness ⍰ = 20, the exponent minimum µ correspond to a concentration  = 0.255, then this concentration is a percolation threshold pc, Different value of the percolation threshold are presented in Table 1

Fig.4 shows the evolution of the width of the fluctuation according the thickness of the system; this width is present in the fig.3.

For a 2d system the fluctuation width equal to 0.15 which proves that there is less fluctuation in the intensity of strong current, for a thickness equal to 3. The fluctuation range expands and the fluctuation width takes a maximum value.  After the addition of another layer the width of fluctuation decreases exponentially fig.4 which proves that we have close to 3d.  After the tenth layer the fluctuation width is saturated so we have a 3d, this are in agreement with those found by Zekri and al [15]

**Conclusion:**

We studied the behavior the width of fluctuation of the strong current for systems between 2d and 3D. We used a method based on the exact resolution of Kirchhoff's equations, this method used to calculate the current intensity of a finished system and variable thicknesses between 2d and 3D

The distributions of the intensity of the currents are interpreted in term of Levy distributions. It was shown that the threshold of percolation pc corresponds at least of the exponent µ who corresponds to the maximum of fluctuations of the intensity of current.
The results obtained show a strong reduction in the width of fluctuation of the strong current as soon as the thickness of the system starts to increase which confirms the result found by Zekri et al [ 15]

-

-

## Table:

**Table.1:** the threshold percolation according the thickness system for the size 100

## Figure:

**Fig1:** a network 3×3×2 with conductivities metal and of insulator distributed by chance

**Fig.2:** distribution of logarithm of current for size 100 and thickness 15 for P= 0.24 and 0.25. The inserted figure watches same distributions traced in log-log.

**Fig. 3:** exponent µ according to the concentration for size 100 and various thicknesses

**Fig.4:** the fluctuation width according to the system thicknesses for size 100. The inset shows the log-log plot of the same figure

-

Tables:

| ⏹ | 3 | 4 | 5 | 8 | 10 | 12 | 15 | 20 | 25 |
|---|---|---|---|---|---|---|---|---|---|
|  | 0.34 | 0.305 | 0.29 | 0.27 | 0.265 | 0.26 | 0.26 | 0.255 | 0.255 |

**Table.1:** the threshold percolation according the system thicknesses of size 100

**Figures:**

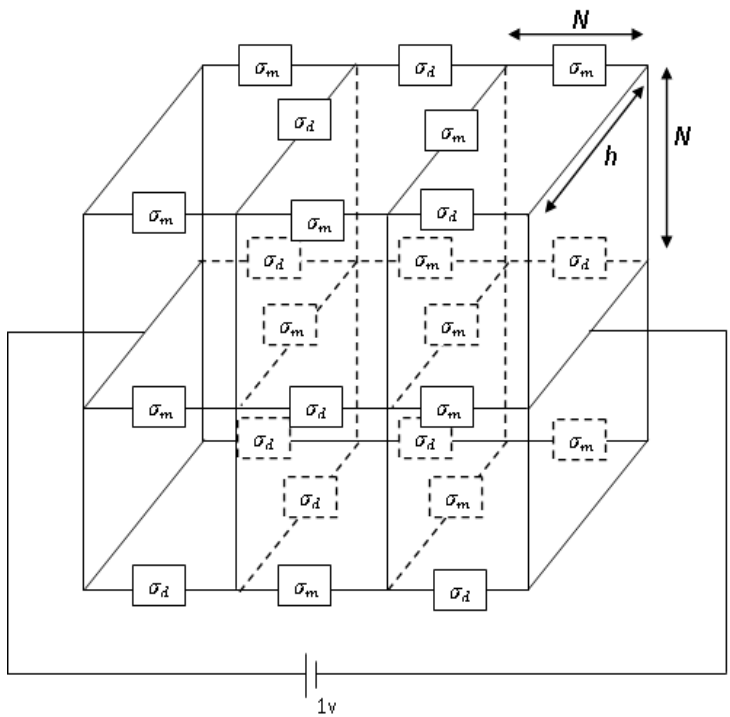

**Fig1:** a network 3×3×2 with conductivities metal and of insulator distributed by chance

-

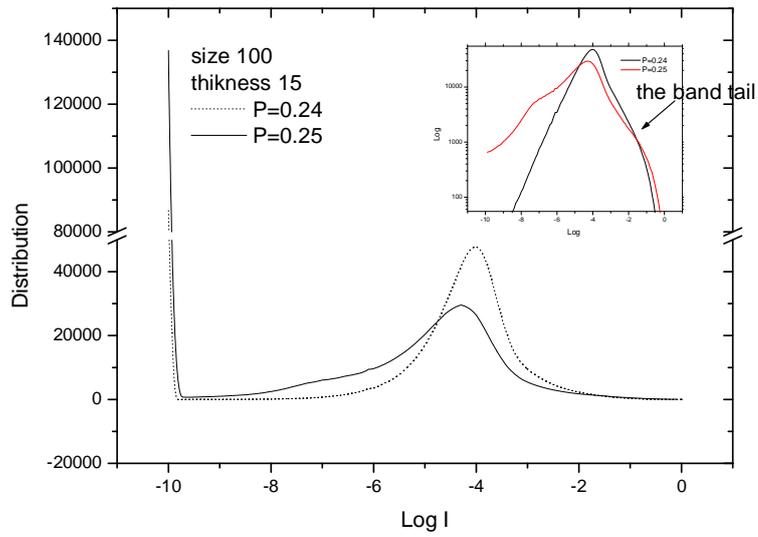

**Fig.2:** distribution of logarithm of current for size 100 and thickness *h=15* for P= 0.24 and 0.25. The inserted figure shows same distributions traced in log-log.

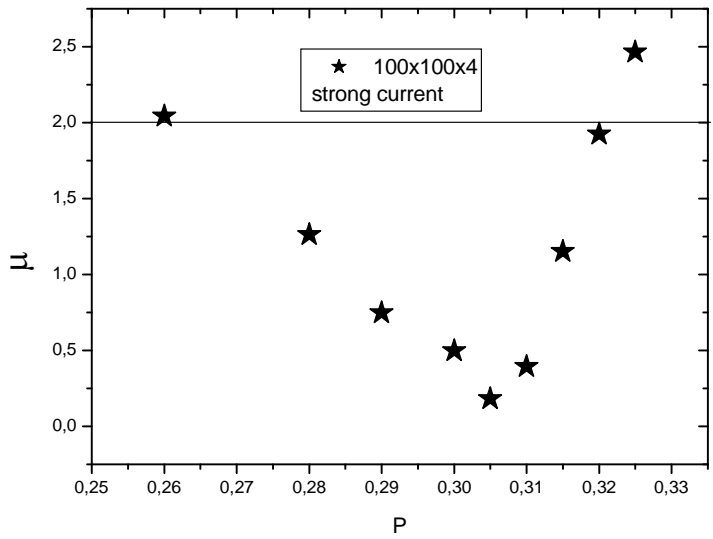

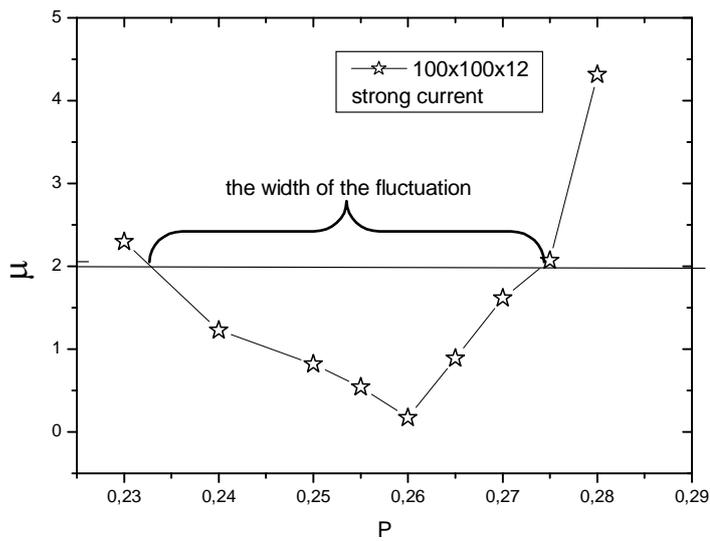

**Fig. 3:** exponent μ according to the concentration for size 100 and various thicknesses

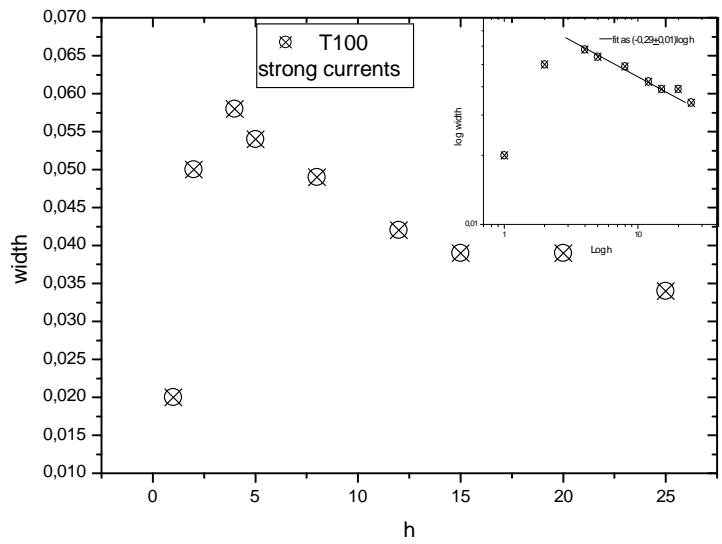

**Fig.4**: the width fluctuation according to system's thickness for size 100. 100. The inset shows the log-log plot of the same figure.